\documentclass{ifacconf}
\usepackage{graphicx}     
\usepackage{natbib}  
\usepackage{color,theorem}

\newtheorem{remark}{Remark}

\usepackage{amssymb}
\usepackage{amsmath}
\usepackage{mathrsfs}
\usepackage{algorithm}
\usepackage{algorithmic}
\usepackage{bm}
\usepackage{xcolor}
\usepackage{color}
\usepackage{pgfplots}
\usepackage{theorem}
\usepackage{enumitem}
\setitemize[1]{itemsep=5pt,partopsep=3pt,parsep=\parskip,topsep=5pt}

\newcommand{\imag}{\mathfrak{i}}
\newcommand{\tr}{\top}

\begin{document}
\begin{frontmatter}

\title{Optimal Experiment Design for AC Power Systems Admittance Estimation } 

\author[First]{Xu Du} 
\author[Second]{Alexander Engelmann} 
\author[First]{Yuning Jiang}
\author[Third]{Timm Faulwasser}
\author[First]{Boris Houska}

\address[First]{School of Information Science and Technology, ShanghaiTech University, Shanghai, China  \\
		\{\tt{duxu, jiangyn, borish\}@shanghaitech.edu.cn}}
\address[Second]{Institute for Applied Computer Science, Karlsruhe Institute of Technology, Karlsruhe, Germany \\ 	{\tt {alexander.engelmann}@kit.edu}}
\address[Third]{Institute of Energy Systems, Energy Efficiency and Energy Economics,  TU Dortmund University, Dortmund, Germany \\
	{\tt {timm.faulwasser}@ieee.org}}

\begin{abstract}
The integration of renewables into electrical grids calls for the development of tailored control schemes which in turn require reliable grid models.
In many cases, the grid topology is known but the actual parameters are not exactly known.
This paper proposes a new approach for online parameter estimation in power systems based on optimal experimental design using multiple measurement snapshots.
In contrast to conventional methods, our method computes optimal excitations extracting the maximum information in each estimation step to accelerate convergence.
The performance of the proposed method is illustrated on a case study.
\end{abstract}

\begin{keyword}
Power System Parameter Estimation, Optimal Experiment Design
\end{keyword}

\end{frontmatter}

\section{Introduction}
The safe and reliable operation and control of power systems with a large share of renewables requires reliable grid models.
In many cases topology information is available while the line parameters are unknown or erroneous~(\cite{Abur2004,Kusic2004}). 
This might lead to difficulties in predicting critical situations which in turn can lead to black-outs and to substantial socio-economic costs.
At the same time, shutting down of critical power systems infrastructure to perform identification procedures is usually  no viable option. 
Thus, online algorithms for determining the parameters in electrical power systems are of significant interest. 

In the present paper, we consider the \emph{stationary} AC power system parameter estimation problem, i.e. the problem of estimating line parameters of the AC power flow equations  neglecting transient phenomena.\footnote{We refer to~(\cite{Zhao2019}) for a recent overview on \emph{dynamic} power system  parameter estimation.} More specifically, we propose to tackle the problem via concepts stemming from Optimal Experimental Design (OED).
%
%
%

Classical static power system parameter estimation can roughly be categorized along two axis: first, methods considering only one time instant of measurements versus methods using multiple ones; and second, methods simultaneously estimating states (voltage magnitude and phase angle) \emph{and} the parameters versus methods only considering the parameters or following a sequential state-parameter estimation procedure cf.~(\cite{Abur2004,Monticelli1999,Zarco2000}).
~\cite{Bian2011} and ~\cite{Slutsker1996} consider pure parameter estimation with multiple time samples, whereas \cite{Quintana1988} rely on a single snapshot. This method is extended to  multiple snapshots in~(\cite{VanCutsem1988}).~\cite{Slutsker1996} present a version based on the Kalman filter including past measurements indirectly via the corresponding a posteriori state estimate and the error covariance matrix.
Combined state and parameter estimation is considered in \citep{Liu1992}. 
An approach for combined topology/parameter estimation which seemingly does not to fit in the above categorization was recently presented in~\citep{Deka2016,Park2018}.

Optimal experiment design as described by~\cite{Pukelsheim93,Franceschini2008} appears to have received only little attention in the power system context.
It is so far used in the context of measurement placement only \citep{Li2011}. 
In contrast, OED is frequently used in the context of
parameter estimation for linear and nonlinear dynamic systems,
especially in the context of chemical process system identification \citep{Koerkel2004,Houska2015}. 
\cite{Lemoine2016} introduce a method of using  OED for reducing the degree of freedom of variables in the field of frame material discovery; a method for reducing the volume of high-throughput experiments based on OED was proposed by \cite{Talapatra2018}. 
\cite{Pronzato2008} highlights the strong relations between experimental design and control such as the use of optimal inputs to obtain precise parameter estimation. 

Methods relying on a large number of samples include more information into the estimation process and thus yield a better performance compared to single snapshot techniques.
Hence we focus on multiple snapshot techniques here. 
There exists two approaches on how past measurements are considered:
One simple approach is to incorporate the measurments of all sampling instances in one big estimation problem which grows with the number of observed measurements.
This leads to potentially intractable estimation problems and therefore recursive methods have been developed considering information about past measurements as parameters in the current estimation step.
In turn this leads to real-time algorithms which are able to estimate parameters while the system is running. Prominent examples for these methods are the recursive least-squares method or the Kalman filter \citep{Ljung1999}.

This paper proposes an approach for applying methods from optimal experimental design to the online estimation of the admittance parameters in AC power grids. Section~\ref{sec:ACmodel} recaps the AC power grid model. The main contribution is presented in Section~\ref{sec:OED}. By relying on available measurements only, i.e. voltage and power measurements of the grid, the proposed method can be categorized as a recursive online estimator.
A distinctive feature of our method is that we design optimal generator power profiles—i.e. excitations—such that as much information as possible is extracted in a single estimation step while keeping the power output at the consumer constant. This strategy enables us to perform optimally designed experiments while, at the same time, ensuring that the grid remains completely
functional and delivers a constant power output to the end-users. In Section~\ref{sec:numRes}, we compare our results to a recursive estimator with constant input on a 5-bus benchmark system. Our results indicate that
the proposed method outperforms classical recursive parameter estimation techniques.

\textit{Notation:}  For a given $a \in \mathbb{R}^n$ and $\mathcal{C}\subseteq\{1,...,n\}$, ${(a_k)_{k \in \mathcal C}\in\mathbb{R}^{|\mathcal{C}|}}$ stacks all $k \in \mathcal C$ elements. Similarly, for a given $A \in \mathbb R^{n \times l}$ and $\mathcal S \subseteq \{ 1, \ldots, n \} \times \{ 1, \ldots, l \}$, $(A_{i,k})_{(i,k) \in \mathcal S}\in\mathbb{R}^{|\mathcal{S}|}$ denotes a vector stacking elements $A_{i,k}$ for all $(i,k)\in\mathcal{S}$. Moreover, $\imag = \sqrt{-1}$ denotes the imaginary unit, and hence $z = \mathrm{Re}(z) +\imag\cdot \mathrm{Im}(z)$.

\section{AC Power Grid Model} \label{sec:ACmodel}
\begin{figure}[htbp!]
	\centering
	\includegraphics[width=0.9\linewidth]{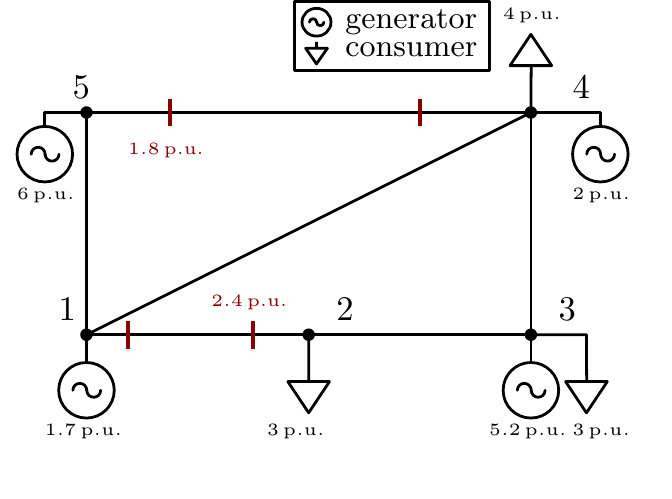}
	\caption{Modified 5-bus system from \cite{Li2010} with $4$ generators and $3$ consumers.}
	\label{fig:ieee5bus}
\end{figure}

Let $(\mathcal{N}, \mathcal{L},Y)$ be a power grid with $\mathcal{N}=\{1,\dots,N\}$ denoting the set of buses, $\mathcal{L}\subseteq \mathcal{N}\times \mathcal{N}$ is the set of transmission lines, and $Y \in \mathbb{C}^{N\times N}$ is the sparse, complex-valued admittance matrix. The admittance matrix is defined as
\[
Y_{k,l}=\left\{
\begin{array}{ll}
\sum\limits_{j \neq k} \left( g_{k,j} + \imag \, b_{k,j} \right) & \text{if} \; k=l, \\  [0.25cm]
- \left( g_{k,l} + \imag \, b_{k,l} \right) & \text{if} \; k \neq l,
\end{array}
\right.
\]
where $g_{k,l}$ are the line conductances and $b_{k,l}$ are the line susceptances for all transmission lines $(k,l) \in \mathcal L$. In general, not all buses are connected.
 Thus,  for most real networks  the matrix $Y$ can be expected to be sparse. Hence we set $g_{k,l} = b_{k,l} =0$ for all $(k,l) \notin \mathcal L$. For example, for the $5$-bus network in Figure~\ref{fig:ieee5bus} the nodes $3$ and $5$ are not directly connected. 
\footnote{ We use the per-unit system which is a standard normalization procedure from power systems \citep{Glover(2012)}.
The base-power is  $100\, \mathrm{MVA}$ and the base-voltage is $230\, \mathrm{kV}$. 
The base-frequency is unfortunately not given in the dataset \citep{Li2010}, but it is also not needed as conductances and susceptances are given.}

Let $v_k$ denote the voltage amplitude at the $k$-th node and $\theta_{k}$ the voltage angle. Throughout this paper, we assume that the voltage magnitude and the voltage angle at the first node, 
\[
\theta_1 = 0 \quad \text{and} \quad v_1 = \mathrm{const.} \; ,
\]
are fixed. This assumption can always be made without loss of generality, since the power flow in the network depends on the relative voltage angles $\theta_k - \theta_l$. Similarly, the voltage $v_1$ at the first node is regarded as the reference voltage. Thus, because $\theta_1$ and $v_1$ are given,  the vector
\[
x = \left( v_2, \theta_2, v_3, \theta_3, \ldots, v_N, \theta_N \right)^\tr
\]
is the state of the power system. In the following, we use the auxiliary function
\begin{align}
P_k(x,y) =& v_k^2 \sum_{l \in \mathcal N_k}  \left(
\begin{array}{rr}
g_{k,l} \\[0.16cm]
-b_{k,l}
\end{array}
\right) \notag \\
& - v_k \sum_{l \in \mathcal N_k} v_l
\left(
\begin{array}{rr}
g_{k,l}  & b_{k,l} \\[0.16cm]
-b_{k,l} & g_{k,l}
\end{array}
\right)
\left(
\begin{array}{c}
\cos(\theta_{k}-\theta_{l}) \\[0.16cm]
\sin(\theta_{k}-\theta_{l})
\end{array}
\right) \notag
\end{align}
in order to model the active and reactive power residuum at the $k$-th node, where the shorthand
$$\mathcal N_k = \{ l \in \mathcal N \, \mid \, (k,l) \in \mathcal L \, \}$$
denotes the set of neighbors of the $k$-th node. Moreover, 
\begin{equation}\label{eq:y}
y = \left(
\begin{array}{c}
g_{k,l} \\[0.16cm]
b_{k,l} 
\end{array}
\right)_{(k,l)\in \mathcal{L}} \in \mathbb{R}^{2|\mathcal{L}|} \notag
\end{equation}
denotes the admittance parameter vector, that is, a vector consisting of all non-zero components of the admittance matrix  $Y$ that are off-diagonal. Notice that  $\dim y = 2|\mathcal{L}|$ grows with the number of transmission lines.

The variables $p^d_k$ and $q^d_k$ denote the active and reactive power demands, which are, for the sake of simplicity, assumed to be known and constant. We set $p^d_k = 0$ and $q^d_k = 0$, if there is no consumer at the $k$-th node. Moreover,  $\mathcal G \subseteq \mathcal N$ is the set of generators in the system. The associated generator active and reactive power at the $k$-th node, with $k \in \mathcal{G}$, are denoted by $p^\mathrm{g}_k$ and $q^\mathrm{g}_k$, respectively. Notice that the net active and reactive net power supply at  Node $k$ are
\[
S_k(u) = \left(
\begin{array}{c}
p_k^\mathrm{g} - p_k^\mathrm{d} \\[0.16cm]
q_k^\mathrm{g} - q_k^\mathrm{d}
\end{array}
\right) \quad \text{if} \quad k \in \mathcal G
\]
and
\[
S_k(u) = \left(
\begin{array}{c}
- p_k^\mathrm{d} \\[0.16cm]
- q_k^\mathrm{d}
\end{array}
\right)  \quad \text{if} \quad k \notin \mathcal G \; .
\]
In the context of this paper, we regard the active and reactive powers at all but the first generator
\begin{align}
\notag
u= \left(
\begin{array}{c}
p_k^\mathrm{g} \\[0.16cm]
q_k^\mathrm{g}\\
\end{array}
\right)_{k \in \mathcal G \setminus \{ 1 \} } \; 
\end{align}
as an input that the grid operator can choose. Notice that the conservation of energy must hold at all nodes, which implies that the power residuum at the nodes must be equal to the supplied power,
\begin{align} \label{eq:PFE}
\forall k \in \mathcal N, \qquad  P_k(x,y) = S_k(u) \; .
\end{align}
In the literature equations \eqref{eq:PFE} are known under the name power-flow equations~\citep{Monticelli1999}. At this point, it is important to be aware of the fact that, because the consumer demand is assumed to be constant and given, the grid operator can choose $u$ but needs to make sure that the active and reactive power at the first generator satisfy
\begin{align}
\label{eq::energy}
\left(
\begin{array}{c}
p_1^\mathrm{g} \\[0.16cm]
q_1^\mathrm{g}\\
\end{array}
\right) = P_1(x,y) + \left(
\begin{array}{c}
p_1^\mathrm{d} \\[0.16cm]
q_1^\mathrm{d}\\
\end{array}
\right),
\end{align}
i.e. the overall power balance holds.
In the power systems literature, Node $1$ is commonly called the \emph{slack} node \citep{Grainger1994}.
Typically, a large generator is connected to this node ensuring that the above power balance can always be satisfied.

\begin{remark}[Considering energy storage]
\textit{Equation~\eqref{eq::energy} can alternatively be satisfied by installing a battery or other storage devices at the first node, which supplies the active and reactive power $p_1^\mathrm{g}$ and $q_1^\mathrm{g}$. The advantage of introducing a storage device is that~\eqref{eq::energy} can be satisfied even if the generators temporarily do not match the consumer demand. 
}
\end{remark}

In summary, the power flow equations can be written compactly as
\begin{align}
\label{eq::powerEq}
P(x,y) = S(u) \; ,
\end{align}
with shorthands
\begin{align}
P = [ P_2^\tr, \ldots, P_N^\tr ]^\tr  \quad \text{and} \quad S = [ S_2^\tr, \ldots, S_N^\tr ]^\tr \; . \notag
\end{align}
Notice that 
$$\mathrm{dim}(P) = \mathrm{dim}(x) = 2(N-1) \; .$$
Moreover, the active and reactive power flow in the transmission line $(k,l) \in \mathcal L$ is given by
\begin{multline*}
\Pi_{k,l}(x,y) = v_k^2 \left(
\begin{array}{r}
g_{k,l} \\[0.16cm]
- b_{k,l} 
\end{array}
\right) \\- v_kv_l
\left(
\hspace{-0.05cm}
\begin{array}{rr}
g_{k,l}  & b_{k,l} \\[0.16cm]
-b_{k,l} & g_{k,l} 
\end{array}
\hspace{-0.05cm}
\right)
\hspace{-0.05cm}
\left(
\hspace{-0.05cm}
\begin{array}{c}
\cos(\theta_{k}-\theta_{l}) \\[0.16cm]
\sin(\theta_{k}-\theta_{l})
\end{array}
\hspace{-0.05cm}
\right),
\end{multline*}
which also depends on the non-zero admittance matrix coefficients $y$, $P_k(x,y) = \sum_{l \in \mathcal N_k} \Pi_{k,l}(x,y)$.

\section{Optimal Experiment Design for Admittance Estimation} \label{sec:OED}
Next, we introduce a repeated Optimal Experiment Design (OED) and parameter estimation procedure for estimating the admittance matrix in AC power networks. 

\subsection{Maximum Likelihood Parameter Estimation}
Throughout this paper we assume that the power flow over the transmission lines as well as the system state $x$ can be measured. Therefore, we consider the measurement function $M:\mathbb{R}^{2(N-1)}\times\mathbb{R}^{2|\mathcal L| }\to \mathbb{R}^m$, $m = 2 (|\mathcal L| + N-1)$
\[
M(x,y) = \left[ x^\tr, (\Pi_{k,l}(x,y))_{(k,l) \in \mathcal L}^\tr \right]^\tr \;.
\]
If the associated measurement error has a Gaussian distribution with zero mean and given variance $\Sigma \in \mathbb R^{m \times m}$, $\Sigma \in \mathbb S_{++}^{m}$, the associated maximum likelihood estimation problem for the unknown admittance coefficients $y$ reads
\begin{equation}
\label{eq:MLE}
\begin{split} 
\min_{x,y}\;\;& \frac{1}{2}\|M(x,y)-\eta\|_{\Sigma^{-1}}^2+\frac{1}{2}\|y-y^-\|_{\Sigma^{-1}_0}^2 \\ \quad\mathrm{s.t.}\;\;& P(x,y)=S(u).
\end{split}
\end{equation}
Here, we assume that $y^- \in \mathbb R^{m}$ is a given initial parameter estimate with given variance $\Sigma_0 \in \mathbb{S}_{++}^m$ and $\eta$ are (possibly noisy) measurements associated with $M(x,y)$.

\subsection{Fisher Information}
The power flow equation, $P(x,y) = S(u)$, has in general multiple solutions. For example, this equation is invariant under voltage angle shifts.
However, if the sensitivity matrix\footnote{Conditions under which the matrix $\frac{\partial}{\partial x} P(x,y)$ has full-rank can be found in~\citep{Hauswirth2018}, where linear indendence constraint qualifications for AC power flow problems are discussed in a more general setting.}
\[
\frac{\partial}{\partial x} P(x,y)
\]
has full rank at an optimal solution $(x,y)$ of~\eqref{eq:MLE} for a given $u$, then we can use the implicit function theorem to show that a locally differentiable parametric solution $x^\star(y,u)$ of the equation $P(x,y) = S(u)$ exists. Moreover, this solution satisfies

\begin{align}
\frac{\partial}{\partial y} x^\star(y,u) &=- \left[ \frac{\partial}{\partial x} P(x,y) \right]^{-1} \frac{\partial}{\partial y} P(x,y)
\notag
\end{align}
and, the Fisher information matrix~(\cite{Pukelsheim93}) of the admittance estimation problem~\eqref{eq:MLE} reads
\begin{align}
\label{eq::Fisher}
\mathcal F(x,y,u) = \Sigma_0^{-1} + T(x,y,u) \Sigma^{-1} T(x,y,u)^\tr \, ,
\end{align}
where the shorthand
\begin{align}
	T(x,y,u) =& \frac{\partial}{\partial y} M(x,y) + \frac{\partial}{\partial x} M(x,y) \frac{\partial}{\partial y} x^\star(y,u)\;  \notag
	\end{align}
is used.
The inverse of the Fisher information matrix, $\mathcal F(x,y,u)^{-1}$, can be interpreted as a linear approximation of the variance matrix of the posterior distribution of the parameter $y$~(\cite{Telen2013}). Because this variance depends on the generator power inputs $u$, these inputs can be used to improve the expected quality of the estimate by using an optimal experiment design procedure, as outlined below.

\subsection{Optimal Experiment Design for AC Power Networks}
Next, we develop an OED procedure for improving the accuracy of admittance estimation. Although there are many OED design objectives possible, we focus on the A-design criterion, because the trace of a matrix can be efficiently  evaluated and differentiated without much computational overhead~\citep{Telen2013,Telen2014}). Now, the OED problem at hand reads
\begin{equation}
\label{eq::OED}
\begin{split}
\min_{x,u}&\;\;\mathrm{Tr} \left( \left[ \mathcal F(x,y^{-},u ) \right]^{-1} \right) + \rho\|u-u^-\|_2^2 \\
\text{s.t.}&\;\;
\left\{
\begin{array}{l}
P(x,y^-) = S(u) \\[0.16cm]
\underline u \leq u \leq \overline u \\[0.16cm]
\underline x \leq x \leq \overline x \\[0.16cm]
\underline u_1 \leq P_1(x,y^-) + \left(
\begin{array}{c}
p_1^\mathrm{d} \\[0.16cm]
q_1^\mathrm{d}\\
\end{array}
\right) \leq \overline u_1
\end{array}
\right..
\end{split}
\end{equation}
Here, $y^-$ denotes the current parameter estimate and $u^-$ denotes the old generator set point used for regularization with regularization parameter $\rho \ll 1$.
Moreover, the lower and upper bounds $\underline u,\; \overline u$ are introduced in order to enforce upper and lower bounds on the generator power. Similarly, the lower and upper bounds $\underline x,\; \overline x$ are used to model physical limitations on the voltage magnitude and angle.
The proposed OED approach to power system state estimation is summarized in Algorithm~\ref{alg:OED}. 

\begin{algorithm}[t]
\caption{Optimal Experiment Design for Power System State Estimation}
\textbf{Input:} Initial guess $y^-$ and variance $\Sigma_0^- \succ 0$, a termination tolerance $\epsilon>0$, and an initial generator set-point $u^-$.\\
\textbf{Repeat:}
\begin{itemize}
\item[1)] \textit{Experiment Design:}  
solve the OED problem~\eqref{eq::OED} and denote the optimal solution for the control input by~$u$.

\item[2)] \textit{Collection of Measurements:} set the active and reactive power at the generators to $u$ and take a measurement $\eta$.

\item[3)] \textit{Maximum Likelihood Estimation:} solve the maximum likelihood estimation problem~\eqref{eq:MLE} for given $u$ and denote the optimal solution by $(x,\;y)$.

\item[4)] \textit{Termination Check:} If the trace of the variance is sufficiently small, $\mathrm{Tr}\left( \left[ \mathcal{F}(x,y,u) \right]^{-1} \right) < \epsilon$, break and return the parameter estimate $y$ as well as its approximate variance $\left[ \mathcal{F}(x,y,u) \right]^{-1}$.

\item[5)] \textit{Update Step:} Set $y^- \leftarrow y$, $\Sigma_0 \leftarrow \left[\mathcal{F}(x,y,u) \right]^{-1}$. Moreover, we set $u^- \leftarrow u$ and return to Step 1).

\end{itemize}
\label{alg:OED}
\end{algorithm}
\begin{figure}[h]
	\centering
	\includegraphics[width=8.7cm]{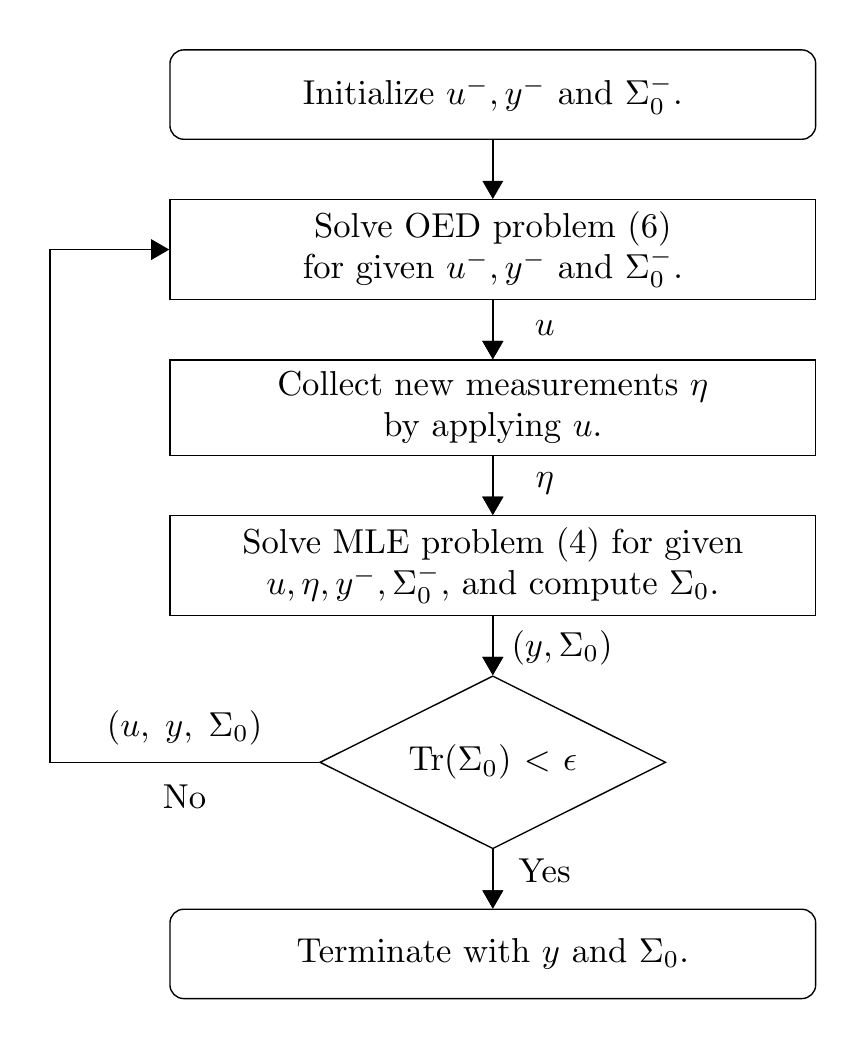}
	\caption{Flow chart of Algorithm~\ref{alg:OED}. }
	\label{flow chart}
\end{figure}

\begin{remark}[Termination in finitely many steps]~\\
Because we assume measurements of the power flow over all transmission lines, it is clear that all admittance coefficients are observable. In other words, the matrix $T(x,y,u)$ always has full-rank---independent of how one chooses $u$. Thus, it follows from~\eqref{eq::Fisher} and from the update of $\Sigma_0$ in Step 5) of Algorithm~\ref{alg:OED} that the Fisher information is strictly monotonically increasing, which, in turn, implies that Algorithm~\eqref{alg:OED} terminates after a finite number of iterations.
\end{remark}

\begin{remark}[Ensuring power balance]
Note that though the algorithm varies the active and reactive power setpoints $p_k^g$ and $q_k^g$, the consumers are not affected by the proposed estimation method as power balance is enforced via the power flow equations \eqref{eq::powerEq}.
\end{remark}

\section{Numerical Results} \label{sec:numRes}
Next we illustrate the performance of Algorithm~\ref{alg:OED} drawing upon the modified 5-bus system from~\citep{Li2010} shown in Figure~\ref{fig:ieee5bus}.

\subsection{Implementation and data}
The problem data is obtained from the \texttt{MATPOWER} dataset ~\citep{Zimmerman2011}.
As discussed in Section 2, we use bus~1 as reference with fixed $v_0$ and $\theta_0$.

\begin{figure}[h]
	\centering
	\includegraphics[width=1\linewidth]{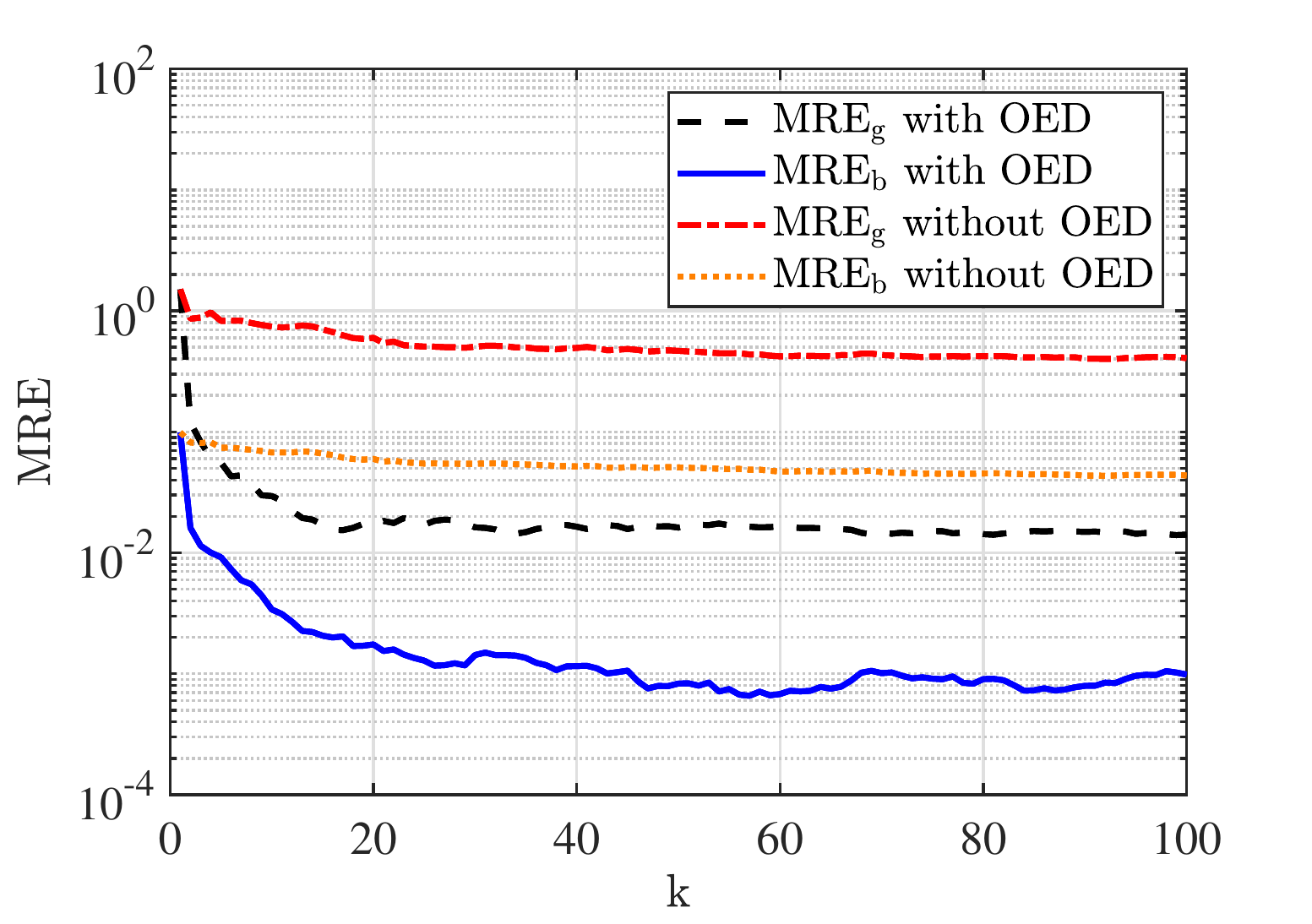}
	\caption{Mean relative errors $\mathrm{MRE}_\mathrm{b}$ (solid line) and $\mathrm{MRE}_\mathrm{g}$ (dashed line) as obtained by Algorithm~\ref{alg:OED}. 
		The dotted line and dash-dotted correspond to the corresponding
		mean relative errors that are obtained by running the estimation without optimally exciting the active and reactive power at the generators. }
	\label{fig:error}
\end{figure}

We consider measurements $M( x(\bar y,u^\star), \bar y) + \chi$ with additive Gaussian noise $\chi$, which has zero mean and covariance $10^{-4}\, I$. 
Here, $\bar y$ denotes the ground truth of $y$ (i.e. the line parameters from the \texttt{MATPOWER} dataset)  and $I$ is the identity matrix. 
Furthermore, we use  $\rho=8\cdot10^{-4}$ as regularization parameter in Step~1 of Algorithm~\ref{alg:OED}. 
We initialize $y^-$ with  a non-zero vector with small norm to avoid numerical difficulties.
Apart from evaluating the total variance of the OED estimator $\mathrm{Tr}(V)$, we also compute the mean relative error of the estimated parameters $y$ via
\begin{align*}
\mathrm{MRE}_\mathrm{g}=\frac{1}{|\mathcal{L}|}\sum_{(k,l)\in \mathcal{L}}\frac{|g_{k,l}-\bar g_{k,l}|}{|\bar g_{k,l}|} \;,\\
\mathrm{MRE}_\mathrm{b}=\frac{1}{|\mathcal{L}|}\sum_{(k,l)\in \mathcal{L}}\frac{|b_{k,l}-\bar b_{k,l}|}{|\bar b_{k,l}|}\;.
\end{align*}
Figure~\ref{fig:error} shows the mean relative error $\mathrm{MRE}_\mathrm{g}$ and $\mathrm{MRE}_\mathrm{b}$ over the iteration index $k$ for two different methods:
For OED and for OED with a constant input $u$ generated in the first iteration of Algorithm~\ref{alg:OED}.

The latter approach is similar to a standard recursive least-squares method. 
One can see that with the one-shot estimate after the first iterate, the relative error is around  $80\%$.
The error can be reduced to a level of around $1\%$ after a couple of iterations.
One can see that the performance of using an optimal input yields a considerably better performance.
Moreover, the strong decrease in $\mathrm{MRE}_\mathrm{g}$ and $\mathrm{MRE}_\mathrm{b}$ after several iterations in both methods underlines the importance of techniques using multiple snapshots.



\begin{table}[htbp!]
	\centering
	\renewcommand{\arraystretch}{1.5}		
	\caption{Line parameter estimation results $[\mathrm{S}]$.} 
	\begin{tabular}{r||p{1.4cm}p{1.4cm}p{1.4cm}p{1.4cm}} 
		\hline
		Index&Conductance & Conductance  & Susceptance &Susceptance \\
		& true value & estimate & true value & estimate \\
		\hline
		$(1,2)$&3.523 & 3.515 &-35.235&-35.233 \\ 
		$(1,4)$&3.257 & 3.274& -32.569&-32.546\\
		$(1,5)$&15.470 &15.364&-154.703&-154.832\\
		$(2,3)$&9.168 & 9.725&-91.676&-91.023\\
		$(3,4)$&3.334 & 3.319&-33.337 &-33.347\\
		$(4,5)$&3.334 &3.351& -33.337  &-33.329\\
		\hline
	\end{tabular}
	\label{table:result}
\end{table}
Table~\ref{table:result} shows the ground truth $\bar y$ and the OED estimation result after 100 iterations. 
One can see that in all cases the relative error is below $6\%$, the $\mathrm{MRE}_\mathrm{g}$ is $1.41\%$ and the $\mathrm{MRE}_\mathrm{b}$ is $0.0985\%$. Notice that the maximum absolute error is $0.557$ Siemens.
\begin{figure}[h]
	\centering
	\includegraphics[width=\linewidth]{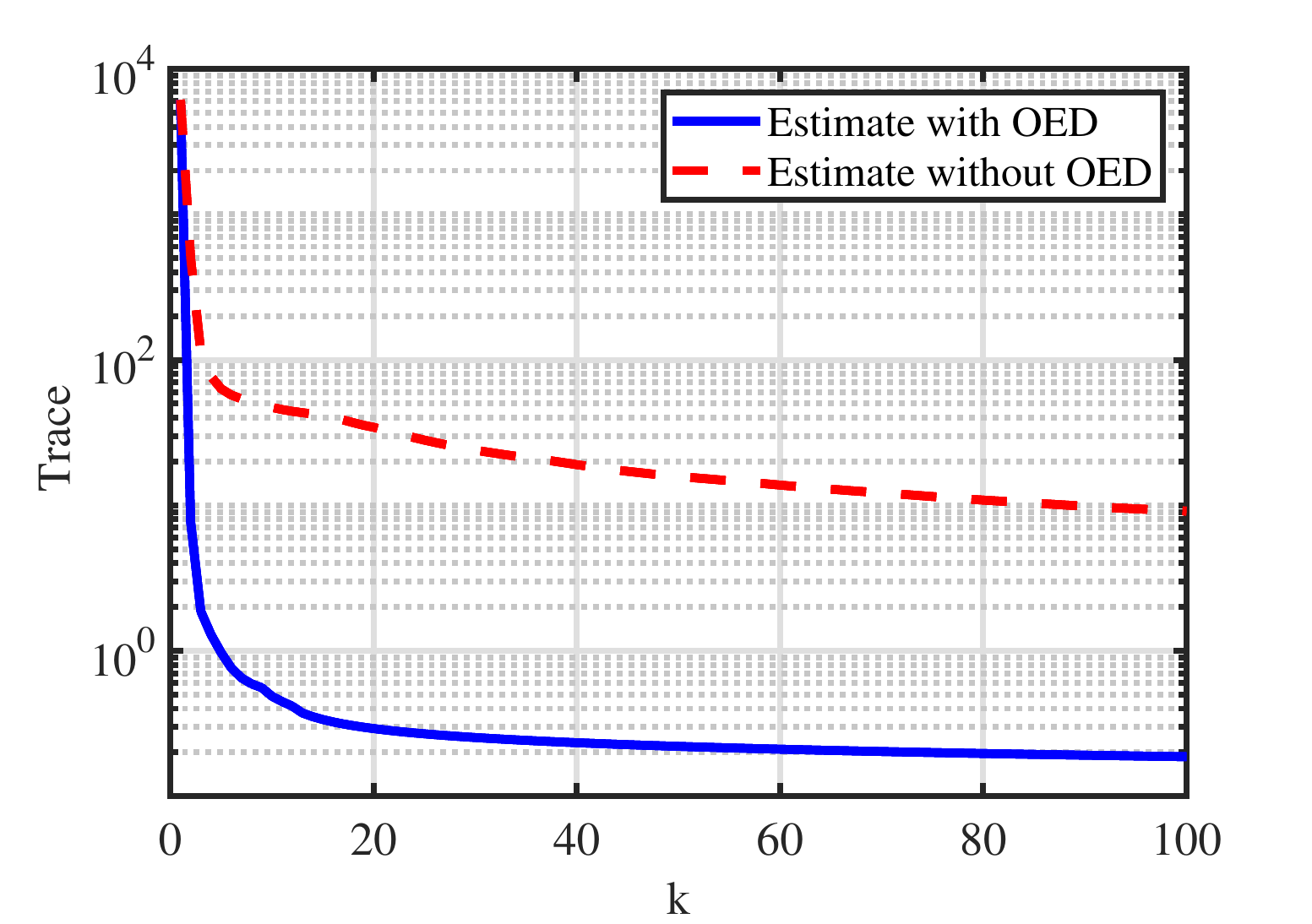}
	\caption{Total variance of the estimation error $\mathrm{Tr}(V(u,y^-))$ obtained by Algorithm~\ref{alg:OED} (solid line) and by running the estimation without optimally exciting the active and reactive power at the generators (dashed line).  }
	\label{fig:trace}
\end{figure}

Figure~\ref{fig:trace} shows $\mathrm{Tr}(V(u,y^-))$ over the iterates $k$.
One can see a monotonic decrease up to a level of $10^{-1}$ within 100 iterations.
The blue solid line represents $\mathrm{Tr}(V(u,y^-))$ using Algorithm~\ref{alg:OED}, and the red dashed  line corresponds to the traditional recursive least square method. 
\begin{figure}[h]
	\centering
	\includegraphics[width=1\linewidth]{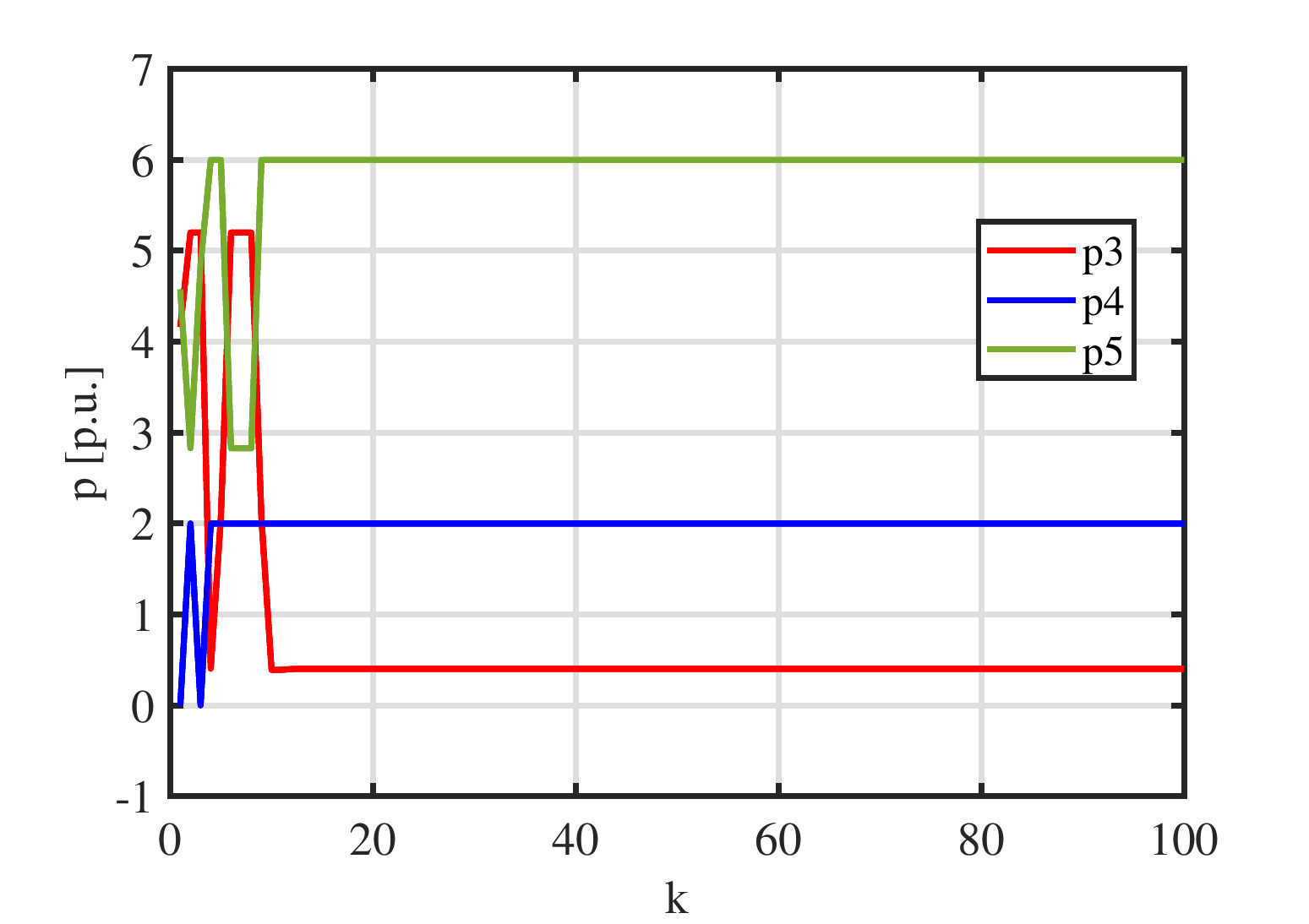}
	\caption{Optimal active power inputs for all generators as obtained by Algorithm~\ref{alg:OED}.}
	\label{fig:active}
\end{figure}

\begin{figure}[h]
	\centering
	\includegraphics[width=1\linewidth]{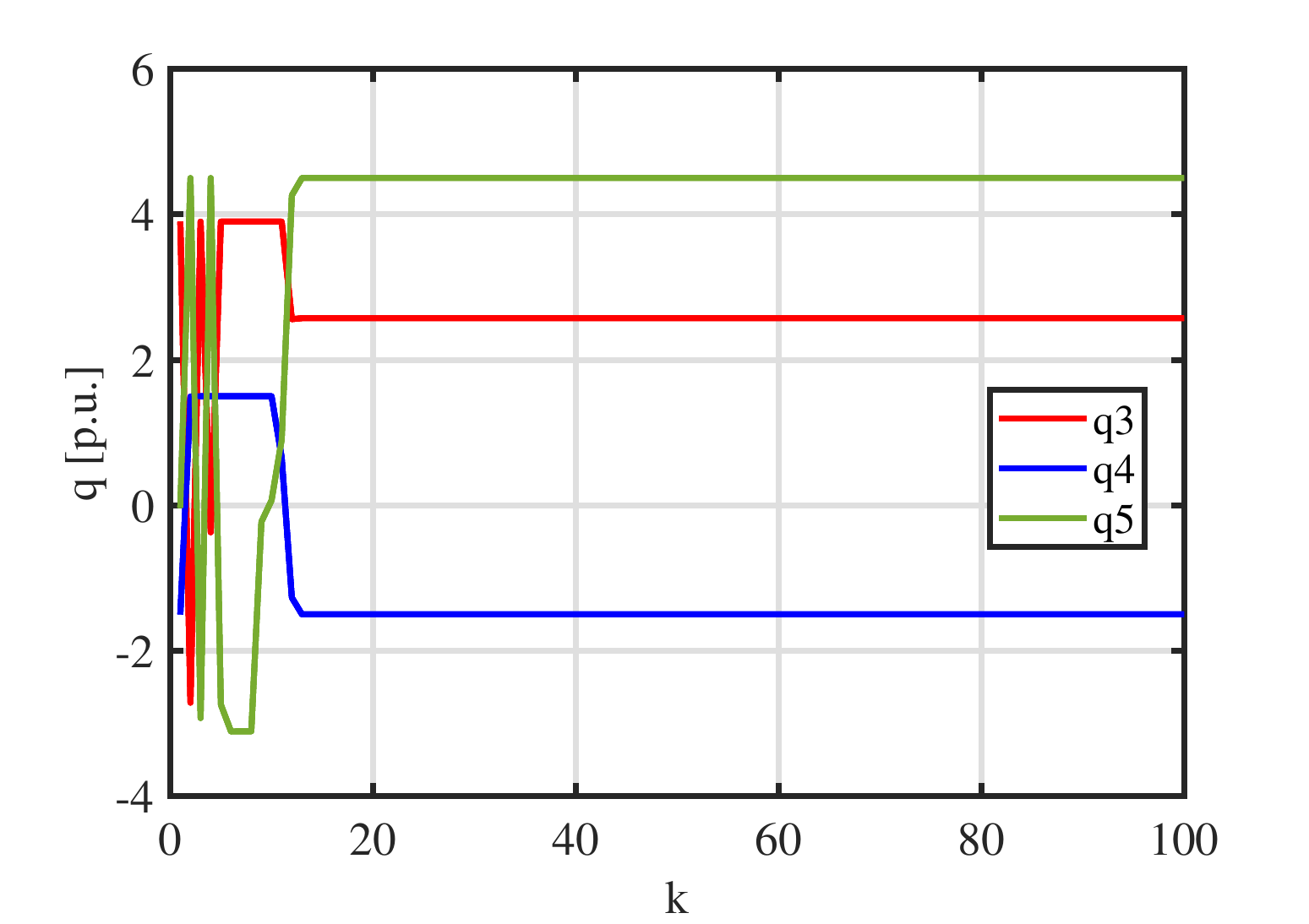}
	\caption{Optimal reactive power inputs for all generators as obtained by Algorithm~\ref{alg:OED}.}
	\label{fig:reactive}
\end{figure}

Figures~\ref{fig:active} and \ref{fig:reactive} show the optimal input for active and reactive power of the three controllable generators in the 5-bus system. 
One can see that after after around 15 iterations, the input remains constant because of the second term in Step 1) of Algorithm~\ref{alg:OED}.
As the Fisher information matrix $\mathcal{F}$ is strictly monotonically increasing its inverse is strictly monotonically decreasing.
With that, the second term in problem~\eqref{eq::OED} starts dominating as Algorithm~\ref{alg:OED} proceeds and thus the change in the optimal input decreases. 


\section{Conclusion}
This work proposed an approach to online power system parameter estimation for the AC grids based on techniques from optimal experimental design. 
Our simulation of a 5-bus AC power system shows that the optimal generator excitation as computed by the proposed method leads to a considerably higher estimation accuracy of the system parameters compared to a recursive least-squares estimation without such excitations.
Specifically, we are able to decrease the mean relative error to less than $1\%$ in a couple of iterations when considering Gaussian measurement noise with a variance of $10^{-4}$ for all the components of the measurement function $M$. 

\section*{ACKNOWLEDGEMENTS}
This work was supported by ShanghaiTech University, Grant-Nr. F-0203-14-012.

 \footnotesize
\bibliography{paper}              
\end{document}